\documentclass{article}
\usepackage{frascatiphys}
\usepackage{graphicx}
\begin{document}
\title{ 
IMPACT OF THE RECOIL SCHEME ON THE ACCURACY OF ANGULAR-ORDERED PARTON SHOWERS
}
\author{
Silvia Ferrario Ravasio, Gavin Bewick     \\
{\em Institute for Particle Physics Phenomenology, Department of Physics, Durham University, Durham} \\
{\em  DH1 3LE, United Kingdom} 
}
\maketitle
\baselineskip=10pt
\begin{abstract}
  In these proceedings we present three possible interpretations of
  the ordering variable implemented in the {\tt Herwig7}
  angular-ordered parton shower. Each interpretation determines a
  different recoil-scheme prescription and we show how it can impact
  the logarithmic accuracy of the algorithm. We also present
  comparisons with LEP data.
\end{abstract}
\baselineskip=14pt
%
\section{Introduction}
General Purpose Monte Carlo (GPMC) generators are fundamental tools
for collider phenomenology, as they are able to simulate fully
realistic collider events, describing both inclusive and exclusive
distributions with high accuracy.  GPMC involve several components.
The event generation starts with the computation of the scattering
process at some hard scale, $Q$, at a fixed order in perturbation
theory (usually at least NLO QCD).  The event is then fed to a Parton
Shower~(PS) algorithm, which handles the emissions of soft and
collinear partons. The PS thus evolves the system from the hard scale,
$Q$, down to a soft scale, $\Lambda$.  At this point we enter the
non-perturbative regime: QCD interactions are so strong that the 
coloured partons are forced to form colour
singlets, \emph{i.e.} they hadronize. To properly simulate hadron
colliders, we also need to provide a model of the
\emph{underlying event}, \emph{i.e.} secondary interactions between 
initial-state partons that do not participate in the hard interaction.

In these proceedings we will focus on the PS component, which
provides a bridge between the perturbative and non-perturbative
regimes of QCD and allows the multiplicity of particles in the event
to increase, which is a key requirement for performing realistic
simulations of collider data.  To achieve this task, the PS exploits
the factorisation properties of QCD in the soft and collinear
limits. When a soft-collinear gluon is emitted from a parton $i$, the
cross section is enhanced and behaves like
\begin{equation}
d \sigma_{n+1} = d \sigma_n \frac{\alpha_s}{\pi} 2 C_i \frac{d \epsilon}{\epsilon}  \frac{d p_T}{p_T} ,
\label{eq:sigmaIR}
\end{equation}
where $d \sigma_n$ is the differential cross-section for the
production of $n$ particles, $C_i$ is the colour factor associated
with the emission from parton $i$ ($C_A$ if $i$ is a gluon, $C_F$ if
it is a quark), $\epsilon$ is the energy fraction carried by the gluon
and $p_T$ is its transverse momentum with respect to the emitter. From
eq.~(\ref{eq:sigmaIR}) we clearly see that we can have two sources of
logarithmic divergence: one associated with $\epsilon\to 0$ and one
with $p_T \to 0$.  Thus, when we generate $m$ emissions we can have at
most $2m$ logarithms: these are the leading logarithms~(LL). It is a
common belief that all of the available PS are able to resum such
logarithms since the splitting kernels that are employed to mimic the emission
of a gluon always approach eq.~(\ref{eq:sigmaIR}) in the
soft-collinear limit.  Many efforts have been made towards reaching
next-to-leading log (NLL) accuracy, \emph{i.e.} $2m-1$ logarithms
for $m$ powers of $\alpha_s$.  For example, the use of quasi-collinear
splitting functions\cite{Catani:2000ef} gives the first subleading
collinear logarithms. If one also adopts the two-loop expression for
the running of $\alpha_s$ and the CMW scheme\cite{Catani:1990rr}, then
all the LL and NLL are included, except for those arising from soft
wide-angle gluon emissions.

Due to the increasing precision of experimental measurements, the
determination of the formal accuracy of a PS is becoming a serious
issue which must be addressed. A recent work\cite{Dasgupta:2018nvj}
introduced an approach to evaluate the logarithmic accuracy based on
the ability of the PS to reproduce the singularity structure of
multi-parton matrix elements, and the logarithmic resummation results.
The authors focus on the process of double gluon emission in $e^+e^-
\to q\bar{q}$ events, where the quark, $q$, is massless and the two
gluons are well separated in rapidity so that the emission probability
reduces to
\begin{equation}
d P_2 = \frac{1}{2}\prod_{i=1}^2 \left(\frac{\alpha_s}{\pi} 2C_F \frac{d p_{T,i}}{p_{T,i}} \frac{d \epsilon_i}{\epsilon_i}\right) = \frac{1}{2}\prod_{i=1}^2 \left(\frac{\alpha_s}{\pi} 2C_F \frac{d p_{T,i}}{p_{T,i}} d y_i \right),
\end{equation}
where $y_i$ is the rapidity of the gluon. The analysis is restricted
to dipole showers, specifically the {\tt
  Pythia}\cite{Sjostrand:2004ef} one, which is the default option of
the {\tt Pythia8}\cite{Sjostrand:2014zea} generator, and the {\tt
  Dire}\cite{Hoche:2015sya} one, available in both {\tt Pythia8} and
{\tt Sherpa}\cite{Gleisberg:2008ta}. The authors identified regions of
phase space where the second gluon emission probability is generated
with the wrong colour factor, namely $C_A/2$ instead of
$C_F$.\footnote{This is a subleading colour issue, as $C_F \to
  C_{A}/2$ in the large-number-of-colours limit.} This happens when
the second gluon, $g_2$, is closest in angle to the first gluon,
$g_1$, in the rest frame of the $q g_1$ (or $\bar{q} g_1$) dipole but
closest to $q$ (or $\bar{q}$) in the original $q\bar{q}$ frame.
Another consequence is that the first gluon must absorb the
transverse-momentum recoil
\begin{equation}
\vec{p}_{T,1} \to \vec{p}_{T,1}-\vec{p}_{T,2}.
\end{equation}
This also breaks the factorisation of the two emissions, as $p_{T,1}$
can vary quite significantly after the generation of another
branching.

Although it is clear that the coherent
formalism\cite{Marchesini:1983bm} implemented in the {\tt
  Herwig7}\cite{Bahr:2008pv} angular-ordered parton shower prevents
the aforementioned subleading colour issue, the impact of the recoil
scheme on the accuracy of the algorithm must be investigated.  In
these proceedings we summarise the findings of
Ref.~\cite{Bewick:2019rbu}, restricting ourselves to the case of a
massless final-state parton shower in $e^+ e^- \to q \bar{q}$ events.

\section{Interpretation of the Ordering Variable}

In this section we present the main features of the {\tt Herwig7}
angular-ordered (final-state) parton shower, focusing on several
possible interpretations of the ordering variable.

\subsection{One Emission}
\begin{figure}
\centering
\qquad \includegraphics[width=0.475\textwidth]{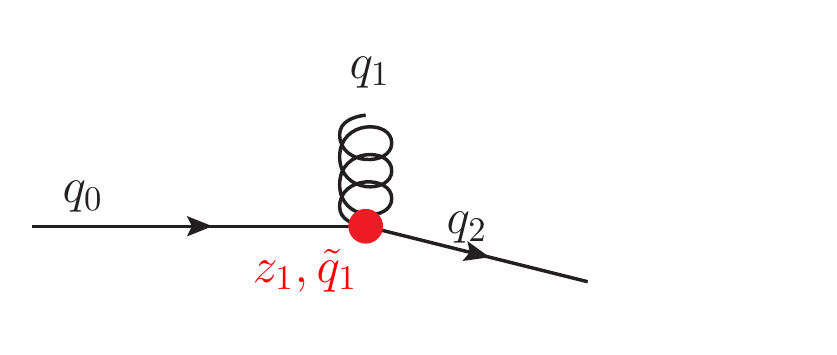}\hspace{-0.2cm} \includegraphics[width=0.475\textwidth]{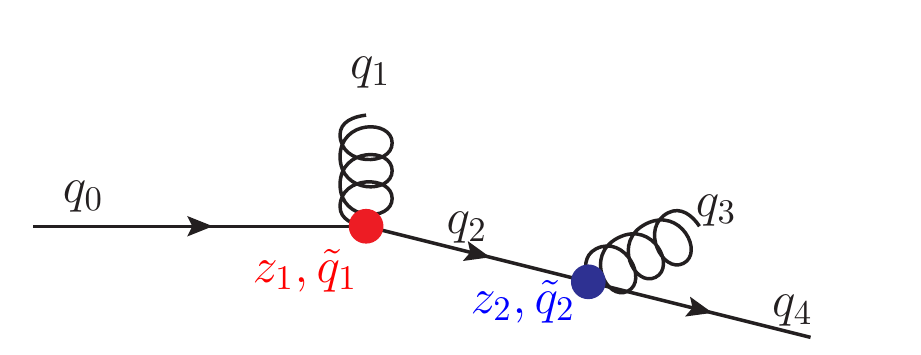}
\caption{Single~(left pane) and double~(right pane) gluon emission from a quark line.}
\label{fig:GluEm}
\end{figure}

We want to generate an emission collinear to the quark, as shown
in the left pane of Fig.~\ref{fig:GluEm}.  We denote with $p$ the
quark momentum and with $n$ a light-like vector parallel to the
momentum of the anti-quark, which is colour connected to the quark in
the original two-body configuration.  The ordering variable can be
equivalently expressed in terms of the transverse momentum
($p_{T,1}$), the virtuality of the emitting quark or the dot-product
of the momenta of the emitted partons,
\begin{equation}
\tilde{q}^2 = \frac{p_{T,1}^2}{z_1^2(1-z_1)^2} = \frac{q_0^2}{z_1(1-z_1)} = \frac{2 q_1 \cdot q_2}{z_1(1-z_1)},
\label{eq:qtildeDef}
\end{equation}
where $z_1$ is the light-cone momentum fraction carried by the emitted
quark.  If we define $\epsilon_1=1-z_1$ we see that in the soft limit,
\emph{i.e.} $\epsilon_1 \to 0$
\begin{equation}
|p_{T,1}| \approx \epsilon_1 \tilde{q}_1, \qquad y_1 \approx -\log \frac{\tilde{q}_1}{Q},
\label{eq:singKin}
\end{equation}
where $Q$ is the centre-of-mass energy, and the {\tt Herwig7} emission
probability approaches the correct limit
\begin{equation}
d P_{\rm Hw7} = \frac{\alpha_s}{2\pi} \frac{d\tilde{q}^2}{\tilde{q}} C_F \frac{1+z_1^2}{1-z_1} d z_1 \to 2 C_F \frac{\alpha_s}{\pi} \frac{d |p_{T,1}|}{|p_{T,1}|} d y_1.
\end{equation}

\subsection{Double Emission} 
We now want to generate the second gluon emission.  If the second
gluon is parallel to the anti-quark, the {\tt Herwig7} algorithm
identifies the $\bar{q}$ as the emitter, the auxiliary vector $n$ is
then chosen to be parallel to the original quark momentum and the
generation of the emission is completely independent to the cascade
originating from the quark.

If both gluons are collinear to the quark, then the requirement that they
have a large rapidity separation suppresses the contribution arising
from the $g \to g g $ splitting and both gluons will be generated
from the quark line with colour factor $C_F$. The angular-ordering
condition $\tilde{q}_2 < z_1 \tilde{q}_1$ dictates that the gluon with 
the smallest rapidity is emitted first, as shown in the right pane of
Fig.~\ref{fig:GluEm}.  The first emitted quark now becomes off-shell
gaining a virtuality $q_2^2 = z_2 (1-z_2) \tilde{q}_2^2$ and the
relations in eq.~(\ref{eq:qtildeDef}) are no longer valid, as it is
impossible to preserve simultaneously $p_{T,1}$, $q_0^2$ and $q_1
\cdot q_2$. The quantity that we preserve determines the recoil-scheme
prescription.

\subsubsection{Transverse-Momentum-Preserving Scheme} 
The original choice\cite{Gieseke:2003rz} was to preserve the
transverse momentum so that we can always write
\begin{equation}
p_{T,i} = z_i (1-z_i) \tilde{q}_i.
\end{equation}
In the soft-collinear limit, the transverse momentum and the rapidity
of each gluon always reproduce eq.~(\ref{eq:singKin}). Thus, two
gluons that are well separated in rapidity are effectively emitted
independently as required.

To preserve the transverse momentum, the virtuality of the previous
emitter must increase
\begin{equation}
q_0^2 = z_1(1-z_1) \tilde{q}_1^2 \to z_1(1-z_1) \tilde{q}_1^2 + \frac{z_2(1-z_2) \tilde{q}_2^2}{z_1}.
\end{equation} 
This tends to produce too much hard radiation in the
non-logarithmically-enhanced region of phase space, overpopulating the
tail of certain distributions.

\subsubsection{Virtuality-Preserving Scheme}
It was then suggested that the virtuality should be
preserved\cite{Reichelt:2017hts}: the transverse momentum of the first
emission is then reduced
\begin{equation}
p_{T,1}^2 = (1-z_1) \left[ z_1^2(1-z_1) \tilde{q}_1^2 - z_2 (1-z_2) \tilde{q}_2^2 \right].
\end{equation}
This choice does not guarantee the existence of a positive
solution. It is easy to see that, even if both emissions are soft, if
the first one is much softer than the second one then there will be a
negative solution, thus breaking the factorisation of multiple gluon
emissions that are well separated in rapidity.

However, it was found that by setting the transverse momentum to 0
whenever a negative solution was encountered, the agreement with the
experimental data is much better than in the $p_T$-preserving scheme.

\subsubsection{Dot-Product-Preserving Scheme}
Motivated by the desire to implement a scheme that is able to
produce independent soft gluon emissions but does not overpopulate
the non-logarithmically-enhanced regions, the last recoil scheme
implemented\cite{Bewick:2019rbu} preserves the dot-product of the
emitted partons and features intermediate properties between the
$p_T$- and $q^2$-preserving schemes.  After $n$ emissions, the
transverse momentum of the first gluon is modified to
\begin{equation}
p_{T,1}^2 = (1-z_1)^2 \left[ z_1^2\tilde{q}_1^2 - \sum_{i=2}^n (1-z_i) \tilde{q}_i^2 \right].
\end{equation} 
Using the angular-ordering condition $z_i \tilde{q}_i >
\tilde{q}_{i+1}$ it can be proven that $p_{T,1}$ cannot became
negative. Furthermore, if all the emissions are soft
\begin{equation}
p_{T,1}^2  \to \epsilon_1^2 (\tilde{q}_1^2 - \sum_{i=2}^n \epsilon_i \tilde{q}_i^2) \approx  \epsilon_1^2 \tilde{q}_1^2,
\end{equation}
i.e. subsequent soft emissions do not affect the transverse momentum
of the previous ones.

The virtuality of the first emitter still increases, however,
\begin{equation}
q_0^2 = z_1 (1-z_1) \tilde{q}_1^2 + \sum_{i=2}^n z_i (1-z_i) \tilde{q}_i^2,
\end{equation}
thus leading, again, to a poor description of the tails of certain
distributions, although better than that provided by the
$p_T$-preserving scheme.

To prevent the virtuality of the original quark and anti-quark from
becoming too large, we can accept the event with a probability given by
\begin{equation}
r = \sqrt{1- 2 \left( \frac{q_q^2+q_{\bar{q}}^{2}}{s} \right) + \left(\frac{q_q^2-q_{\bar{q}}^{2}}{s}\right)^2 },
\label{eq:phsp}
\end{equation}
where $\sqrt{s}$ is the centre-of-mass energy, $q_q^2$ is the
virtuality developed by the quark and $q_{\bar{q}}^{2}$ by the
anti-quark.  The factor $r$ comes from the fact that the original
underlying two-body phase space is reduced when the particles increase
their mass.  With the inclusion of this factor, the phase-space
factorisation becomes exact.  We can easily see that for
soft-collinear emissions $r \to 1$, thus this veto does not affect the
logarithmically-enhanced contributions but it can introduce, at most,
power corrections.

\section{LEP results}
In this section we present the results of our simulations obtained
with the \texttt{Herwig7} generator and compare them with data from
LEP.

\begin{figure}[tbh!]
\includegraphics[width=0.48\textwidth]{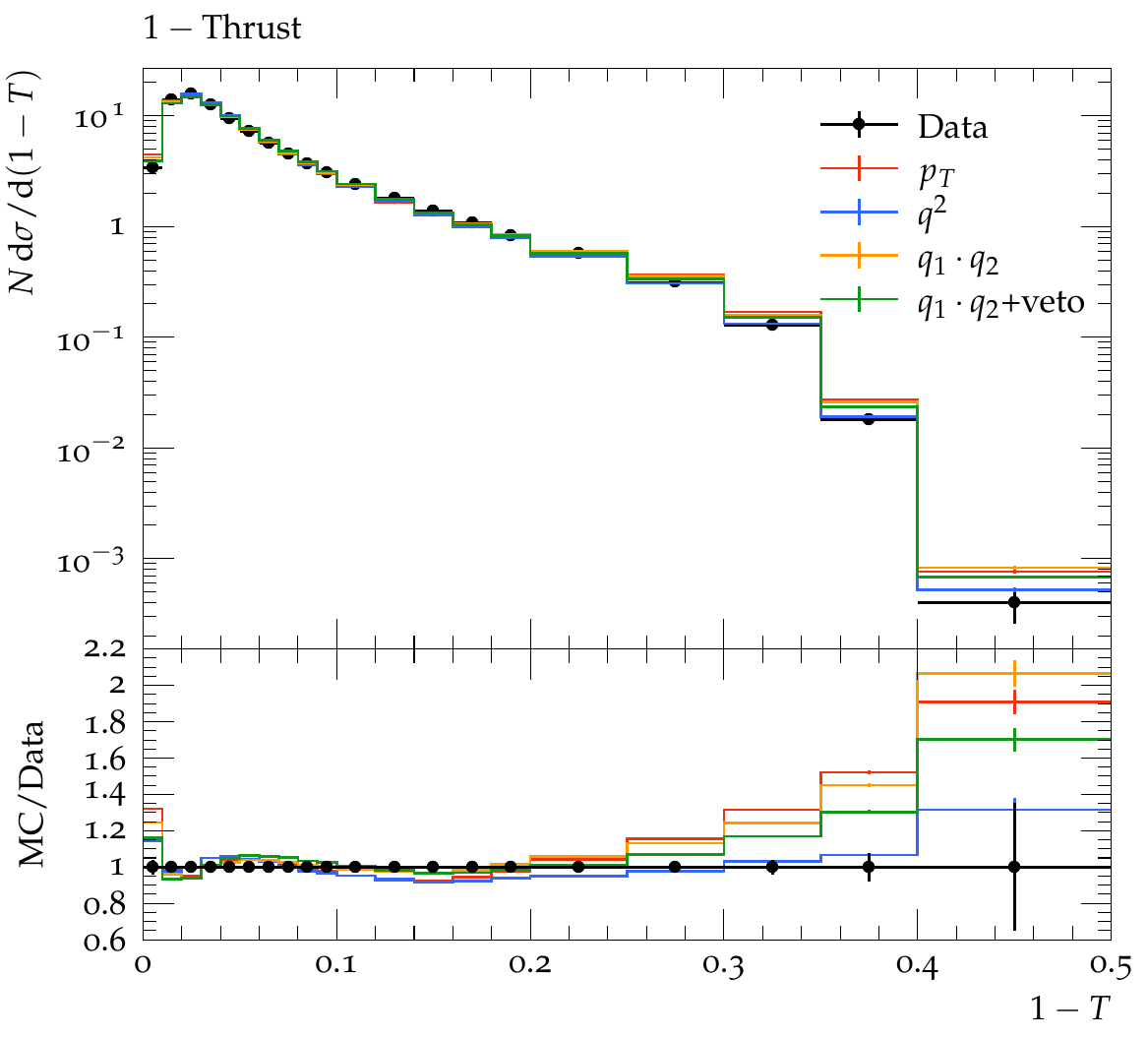} \qquad
\includegraphics[width=0.48\textwidth]{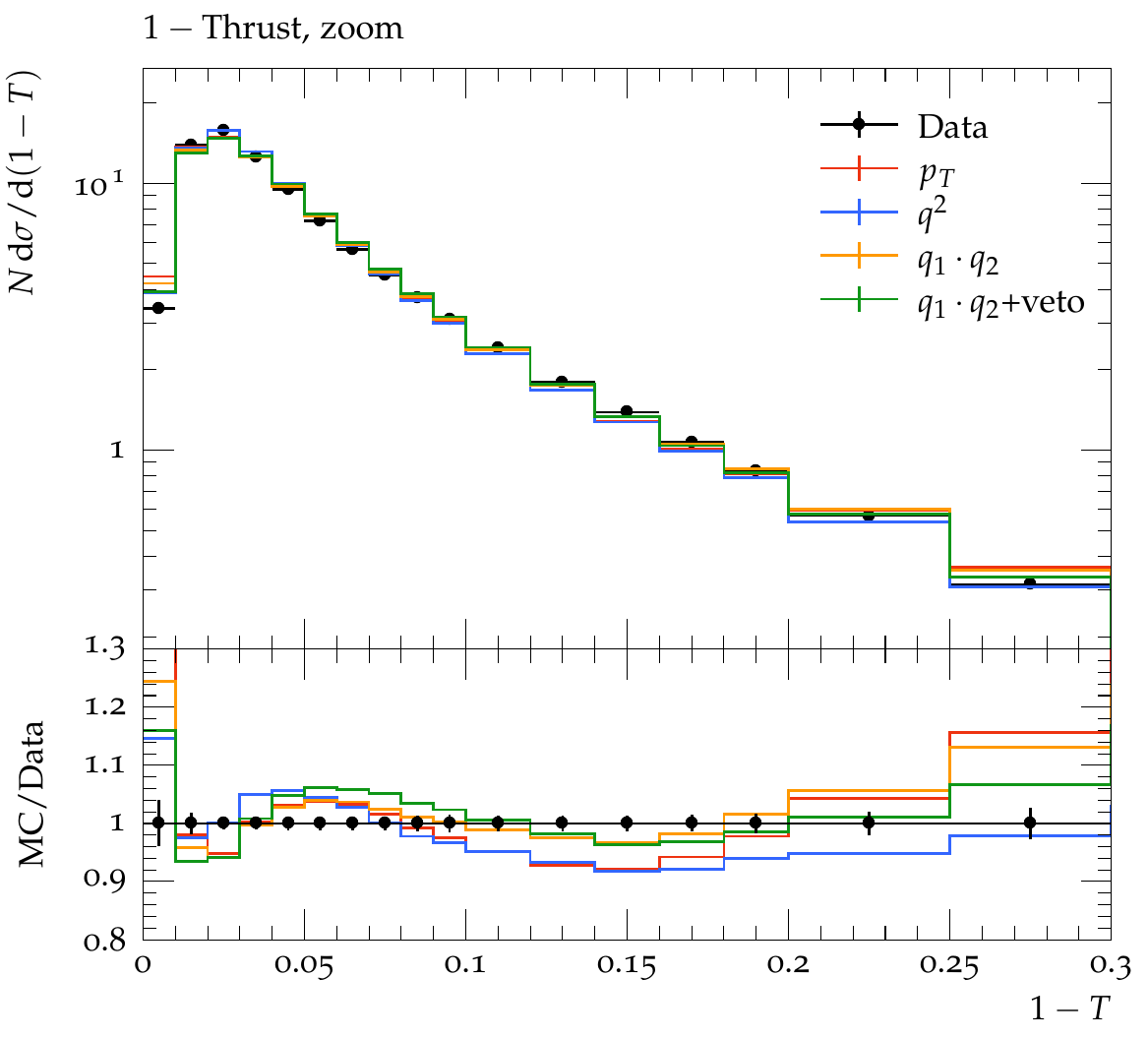}
\caption{The thrust distribution at the Z-pole compared with data from
  the DELPHI\cite{Abreu:1996na} experiment. The right panel gives an
  expanded view of the same for small values $1-T$.}
\label{fig:thrust}
\end{figure}

We begin by showing the thrust distribution (Fig.~\ref{fig:thrust}),
which can be considered as a proxy for all shape distributions.  The
$p_T$- and dot-product preserving schemes overpopulate the tail of the
distribution, which corresponds to the non-logarithmically-enhanced
region of the phase space. Conversely, the $q^2$-preserving scheme
leads to a worse description of the data for $1-T\le 1/3$.  When we
apply the phase space veto (\emph{i.e.} we accept the event with
probability given by eq.~(\ref{eq:phsp})), the behaviour of the
dot-product scheme improves in the tail, leading the best overall
agreement with the data.

\begin{figure}[tbh!]
\includegraphics[width=0.48\textwidth]{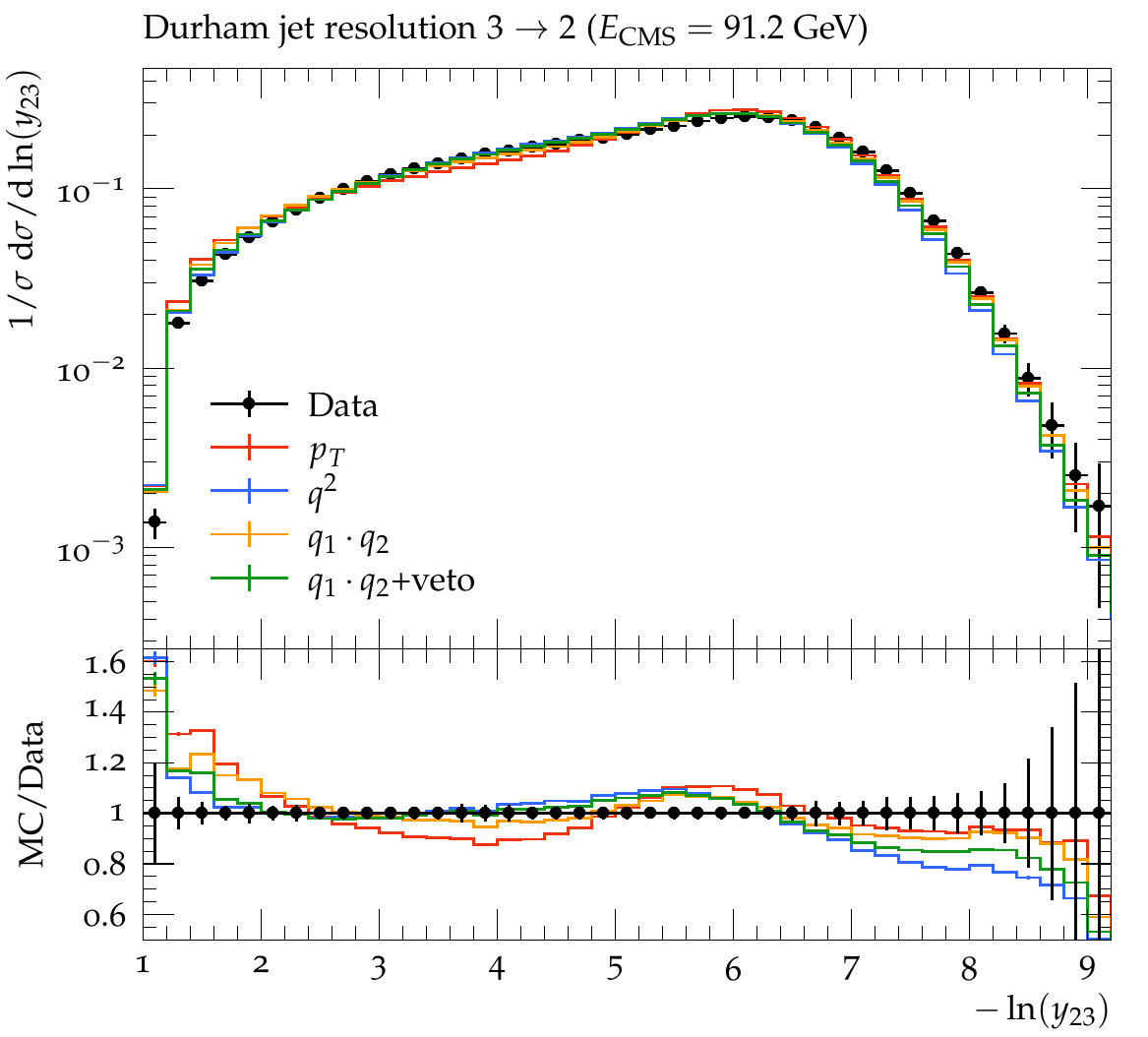} \qquad
\includegraphics[width=0.48\textwidth]{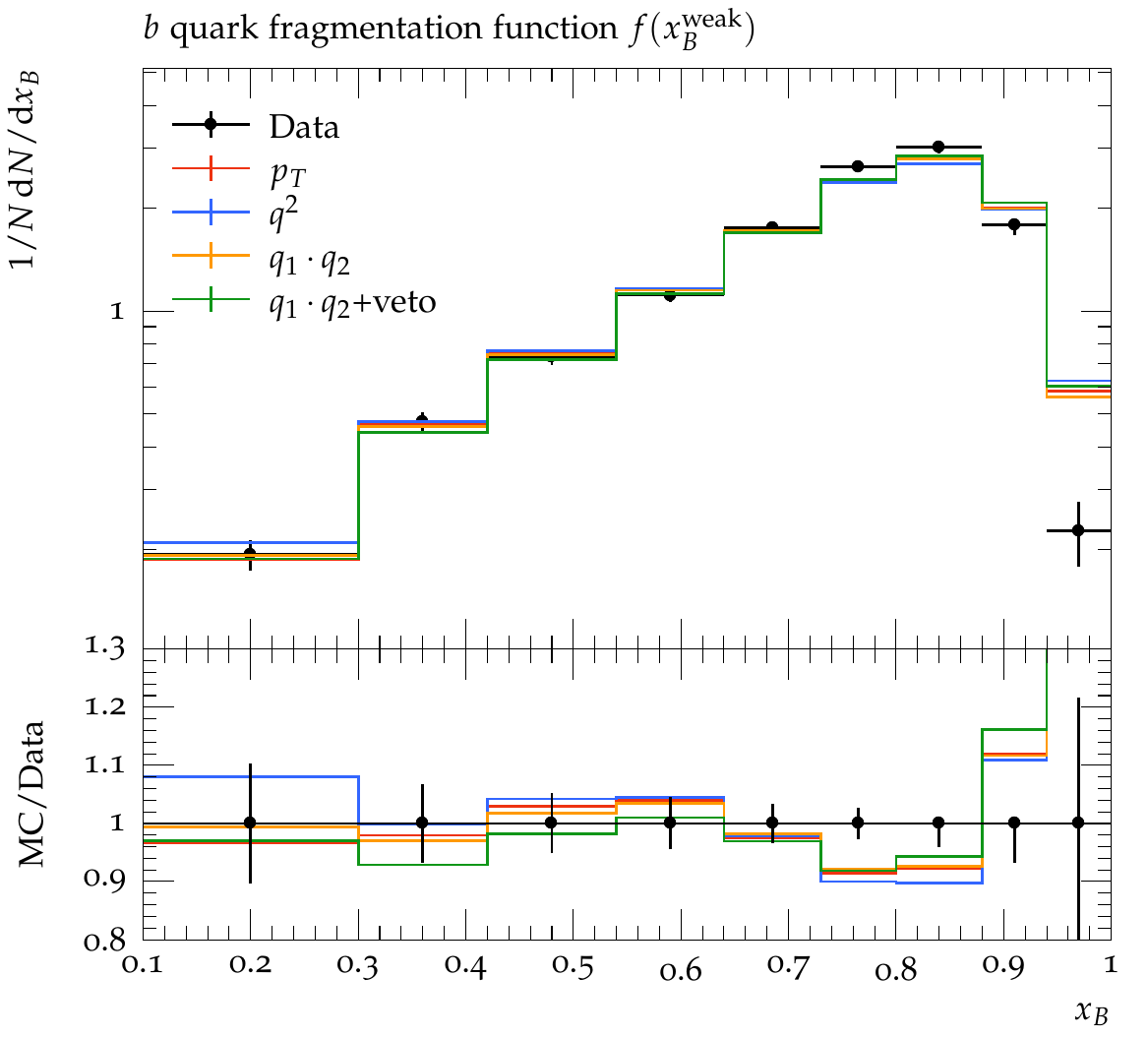}
\caption{In the left panel the 3-to-2 jet resolution parameter for the
  Durham algorithm at the Z-pole compared with data from the
  ALEPH\cite{Heister:2003aj} experiment. In the right panel the
  fragmentation function of weakly-decaying $B$-hadrons compared with
  data from DELPHI\cite{DELPHI:2011aa}.}
\label{fig:y_x}
\end{figure}

The jet-resolution-parameter distribution is shown in the left panel
of Fig.~\ref{fig:y_x}.  As in the previous case, the $q^2$ scheme is
the most accurate in the non-logarithmically-enhanced region (that
corresponds to small values of $-\log(y_{23})$), while the $p_T$
scheme provides the best description in the opposite limit, but gives
the worst overall agreement. The dot-product scheme with the veto is
similar to the $q^2$ scheme, while without the veto it leads to the
best overall agreement with data.

From the right panel of Fig.~\ref{fig:y_x} it can bee seen that none
of the schemes are able to reproduce the bottom quark fragmentation
function for large values of $x_B$.  Issues related to multiple
emissions from heavy quarks, as well as gluons splitting into heavy
quarks, are currently subjects of further investigation.

\section{Summary and Outlook}

Motivated by Ref.~\cite{Dasgupta:2018nvj} we have investigated the
impact that the choice of recoil scheme has on the accuracy of the
\texttt{Herwig7} angular-ordered PS. We found that although the
$p_T$-preserving recoil scheme ensures the independence of successive
soft-collinear emissions well separated in rapidity, it produces too
much radiation in the non-logarithmically-enhanced region of phase
space. The $q^2$-preserving scheme, on the other hand, avoids
overpopulating this region of phase space but breaks the independence
of successive emissions and therefore loses logarithmic accuracy. We
introduced the dot-product-preserving scheme as an attempt to retain
the best features of both schemes, but it still somewhat overpopulates
non-logarithmically-enhanced region of phase space. To ameliorate this
we went on to introduce a phase-space veto that suppresses events with
large-virtuality partons. In these proceedings we did not mention the
effect of quark masses, although this is considered to some extent in
Ref.~\cite{Bewick:2019rbu} and is an ongoing area of research we hope
to address further in future publications.

\section{Acknowledgements}
SFR thanks the organisers of LFC19 for the invitation and the
STRONG-2020 network for the financial support.
The authors thank Peter Richardson and Mike Seymour for useful comments on
the manuscript.  This work has received funding from the UK Science
and Technology Facilities Council (grant numbers ST/P000800/1,
ST/P001246/1) and the European Union’s Horizon 2020 research and
innovation programme as part of the Marie Sk\l{}odowska-Curie
Innovative Training Network MCnetITN3 (grant agreement no. 722104). GB
thanks the UK Science and Technology Facilities Council for the award
of a studentship.


\begin{thebibliography}{99}
\bibitem{Catani:2000ef}
  S.~Catani, S.~Dittmaier and Z.~Trocsanyi,
  Phys.\ Lett.\ B {\bf 500} (2001) 149
  doi:10.1016/S0370-2693(01)00065-X
  [hep-ph/0011222].
 
\bibitem{Catani:1990rr}
  S.~Catani, B.~R.~Webber and G.~Marchesini,
  Nucl.\ Phys.\ B {\bf 349} (1991) 635.
  doi:10.1016/0550-3213(91)90390-J

\bibitem{Dasgupta:2018nvj} 
  M.~Dasgupta, F.~A.~Dreyer, K.~Hamilton, P.~F.~Monni and G.~P.~Salam,
  JHEP {\bf 1809}, 033 (2018)
  doi:10.1007/JHEP09(2018)033
  [arXiv:1805.09327 [hep-ph]].

\bibitem{Sjostrand:2004ef}
  T.~Sj\"{o}strand and P.~Z.~Skands,
  Eur.\ Phys.\ J.\ C {\bf 39} (2005) 129
  doi:10.1140/epjc/s2004-02084-y
  [hep-ph/0408302].

\bibitem{Sjostrand:2014zea}
  T.~Sj\"{o}strand {\it et al.},
  Comput.\ Phys.\ Commun.\  {\bf 191} (2015) 159
  doi:10.1016/j.cpc.2015.01.024
  [arXiv:1410.3012 [hep-ph]].

\bibitem{Hoche:2015sya}
  S.~H\"{o}che and S.~Prestel,
  Eur.\ Phys.\ J.\ C {\bf 75} (2015) no.9,  461
  doi:10.1140/epjc/s10052-015-3684-2
  [arXiv:1506.05057 [hep-ph]].

\bibitem{Gleisberg:2008ta} 
  T.~Gleisberg, S.~Hoeche, F.~Krauss, M.~Schonherr, S.~Schumann, F.~Siegert and J.~Winter,
  JHEP {\bf 0902}, 007 (2009)
  doi:10.1088/1126-6708/2009/02/007
  [arXiv:0811.4622 [hep-ph]].

\bibitem{Marchesini:1983bm}
  G.~Marchesini and B.~R.~Webber,
  Nucl.\ Phys.\ B {\bf 238} (1984) 1.
  doi:10.1016/0550-3213(84)90463-2

\bibitem{Bahr:2008pv}
  M.~Bahr {\it et al.},
  Eur.\ Phys.\ J.\ C {\bf 58} (2008) 639
  doi:10.1140/epjc/s10052-008-0798-9
  [arXiv:0803.0883 [hep-ph]].

\bibitem{Bewick:2019rbu} 
  G.~Bewick, S.~Ferrario Ravasio, P.~Richardson and M.~H.~Seymour,
  arXiv:1904.11866 [hep-ph].

\bibitem{Gieseke:2003rz} 
  S.~Gieseke, P.~Stephens and B.~Webber,
  JHEP {\bf 0312}, 045 (2003)
  doi:10.1088/1126-6708/2003/12/045
  [hep-ph/0310083].

\bibitem{Reichelt:2017hts}
  D.~Reichelt, P.~Richardson and A.~Siodmok,
  Eur.\ Phys.\ J.\ C {\bf 77} (2017) no.12,  876
  doi:10.1140/epjc/s10052-017-5374-8
  [arXiv:1708.01491 [hep-ph]].
  
\bibitem{Abreu:1996na}
  P.~Abreu {\it et al.} [DELPHI Collaboration],
  Z.\ Phys.\ C {\bf 73} (1996) 11.
  doi:10.1007/s002880050295

\bibitem{Heister:2003aj}
  A.~Heister {\it et al.} [ALEPH Collaboration],
  Eur.\ Phys.\ J.\ C {\bf 35} (2004) 457.
  doi:10.1140/epjc/s2004-01891-4

\bibitem{DELPHI:2011aa}
  J.~Abdallah {\it et al.} [DELPHI Collaboration],
  Eur.\ Phys.\ J.\ C {\bf 71} (2011) 1557
  doi:10.1140/epjc/s10052-011-1557-x
  [arXiv:1102.4748 [hep-ex]].

\end{thebibliography}
\end{document}